\documentclass[useAMS,usenatbib,referee]{biom}

 \newcommand{\bitem}{\begin{itemize}}
 \newcommand{\eitem}{\end{itemize}}

\usepackage[figuresright]{rotating}
\usepackage{natbib,amsmath,tabularx,multirow,multicol,adjustbox,threeparttable,float,caption,lineno,threeparttable}

\title[Neuroimaging Biomarker Detection in Mediation Models]{A Novel Strategy for Detecting Multiple Mediators in High-Dimensional Mediation Models}

\author{Pei-Shan Yen$^{1}$\email{dbhaumik@uic.edu}, Soumya Sahu$^{1}$, Debarghya Nandi$^{1}$, Zhaoliang Zhou$^{1}$, \\
\textbf{Olusola Ajilore$^{2}$, and Dulal Bhaumik$^{1,2,*}$ } \\
$^{1}$Division of Epidemiology and Biostatistics, University of Illinois at Chicago, Chicago, US \\
$^{2}$Department of Psychiatry, University of Illinois at Chicago, Chicago, US}

\begin{document}
\nolinenumbers




\label{firstpage}

\begin{abstract}
This article presents a novel methodology for detecting multiple biomarkers in high-dimensional mediation models by utilizing a modified Least Absolute Shrinkage and Selection Operator (LASSO) alongside Pathway LASSO. This approach effectively addresses the problem of overestimating direct effects, which can result in the inaccurate identification of mediators with nonzero indirect effects. To mitigate this overestimation and improve the true positive rate for detecting mediators, two constraints on the $L_1$-norm penalty are introduced. The proposed methodology's effectiveness is demonstrated through extensive simulations across various scenarios, highlighting its robustness and reliability under different conditions. Furthermore, a procedure for selecting an optimal threshold for dimension reduction using sure independence screening is introduced, enhancing the accuracy of true biomarker detection and yielding a final model that is both robust and well-suited for real-world applications. To illustrate the practical utility of this methodology, the results are applied to a  study dataset  involving patients with internalizing psychopathology, showcasing its applicability in clinical settings. Overall, this methodology signifies a substantial advancement in biomarker detection within high-dimensional mediation models, offering promising implications for both research and clinical practices.
\end{abstract}


\maketitle
\section{Introduction \label{section intro}}
Understanding the impact of pharmacological interventions on neuroimaging biomarkers is crucial for improving therapeutic outcomes in brain disorders. Identifying neuroimaging biomarkers—such as brain connectivity, structural changes, metabolic activity, and functional responses—offers valuable insights into the intricate interactions within neural networks. Among these biomarkers, brain connectivity is particularly significant, as it reflects the interactions between distinct brain regions within both cortical and subcortical networks. Aberrant changes in brain connectivity can lead to neurological disorders, resulting in dysfunctions in cognitive processes that manifest as abnormal physical and mental behaviors. By assessing these relationships, researchers can develop more effective treatments, facilitate early diagnosis, and tailor interventions, such as neuromodulation for individuals with neurological disorders.

Brain connectivity encompasses effective connectivity (EC), structural connectivity (SC), and functional connectivity (FC). Functional connectivity, in particular, reflects the temporal correlation of neurophysiological events across spatially separated neural assemblies, providing critical insights into the coordination and integration of brain activity during various cognitive processes \citep{logothetis2008we, friston2009modalities, cohen2008defining, greicius2009resting}. Our research aims to elucidate the neural mechanisms of FC that connect therapeutic interventions for improvements in neurobehavioral outcomes. To achieve this, we develop a high-dimensional mediation model that delineates neural processes, with FC measures acting as mediating variables. This approach emphasizes the importance of decomposing treatment effects into distinct pathways, thereby offering a more comprehensive understanding of the underlying neurobiological mechanisms.

High-dimensional mediation models encounter significant challenges in identifying neuroimaging biomarkers, primarily due to issues related to small sample sizes, which lead to unstable parameter estimates and insufficient statistical power \citep{bhaumik2023power}. To mitigate these challenges, various dimension reduction strategies have been extensively investigated. In our pursuit of improved model interpretability and therapeutic efficacy, we focus on regularization methods, particularly the Least Absolute Shrinkage and Selection Operator (LASSO) \citep{tibshirani1996regression}. The LASSO method and Elastic Net (EN) \citep{zou2005regularization}   primarily regularize parameters associated with the total indirect effect (TIE). The Pathway LASSO method \citep{zhao2021multimodal, zhao2022pathway, zhao2022multimodal}, built upon the EN framework, is specifically designed to penalize the TIE and effectively identify mediators along its  pathway, thereby enhancing the robustness of the analysis. 

Implementing LASSO or Pathway LASSO in mediation models poses several challenges related to parameter estimation.  A primary issue is the overestimation of the direct effect (DE) \citep{pieters2017meaningful}. This problem intensifies with an increasing number of mediators ($p$) or a decreasing sample size ($n$). The overestimation arises because when the TIE is not fully accounted for during the estimation process, it contributes to an erroneous estimation of DE, as the total effect (TE) is the sum of DE and TIE. For instance, LASSO focuses on estimating the parameters associated with TIE. Without any explicit constraints on DE, it tends to overestimate DE and underestimate TIE, adversely impacting the model accuracy, especially the true positive rate (TPR) of mediators with nonzero effects. Even with Pathway LASSO, which implements an $L_1$-norm penalty to DE, the problem persists.

The second challenge resides in the difficulty of accurately identifying mediators with nonzero effects, commonly known as true signals, which is essential for understanding the underlying mechanisms at play.  Commonly used  regularization methods, such as LASSO and Pathway LASSO, apply an $L_1$-norm penalty with equal weights to both parameters involved in the indirect effect (IE).  However, the constraint of equal weights places excessive restrictions on one set of  parameters, diminishing the likelihood of detecting small values of those parameters and potentially lowering the true positive rate (TPR) of the IE. Therefore, exploring an alternative approach that employs a constraint with a reduced magnitude for the tuning parameter  warrants further investigation.

The third challenge pertains to ultra-high-dimensional mediators, where the number of mediators significantly exceeds the sample size, leading to substantial computational difficulties and a marked reduction in statistical power. The application of regularization techniques in such models often results in an extremely low true positive rate (TPR) for the indirect effect (IE) due to excessive noise and sparsity issues. To address these challenges, screening procedures like sure independence screening (SIS) \citep{fan2008sure} are frequently employed prior to regularization. However, a systematic approach for determining the appropriate dimensionality in mediation models remains elusive. Traditionally, the reduced dimension is set as $d=k[n/\log(n)]$, with $k=1$ for linear models. In mediation analysis, where the IE involves two parameter vectors, higher values of $k$ (such as 2 or 3) have been suggested to enhance IE detection \citep{zhang2016estimating, perera2022hima2, luo2020high}. Therefore, a methodological approach is needed to optimize the scaling factor $k$, achieving a balance between computational efficiency and the accurate detection of true signals in ultra-high-dimensional scenarios. This optimization is essential for improving the overall performance of mediation analysis in complex, high-dimensional settings.

The final challenge lies in the limited availability of R packages designed for high-dimensional mediator analysis that utilize regularization methods. Most existing R packages primarily focus on transformation techniques, such as those implemented in \texttt{hdmed}, or on Bayesian modeling approaches. To the best of the authors’ knowledge, only one R package, \texttt{ regmed}, is related to regularization methods; however, it is confined to the LASSO penalty. This gap in the statistical landscape highlights the urgent need for the development of R packages that can address the optimization problem using a wider array of penalty strategies. Such advancements would significantly enhance researchers' ability to conduct high-dimensional mediation analyses more effectively and accurately, ultimately contributing to the  growth and innovation in the field.

In response to the aforementioned challenges, this research work is proposing novel strategies for tuning parameters, optimizing SIS procedures, and developing new R packages for  identifying mediators in high-dimensional settings. Our work contributes four notable innovations as follows: 

We develop strategy for improving the accuracy of DE estimation in high-dimensional mediation models through a tuning parameter for the $L_1$-norm penalty. It introduces an additional constraint to enhance the TPR of IE by adjusting the tuning parameter associated with mediation framework parameters. A guideline for selecting an appropriate scaling factor for dimension reduction in ultra-high dimensional mediators is also presented, emphasizing the balance between computational cost and informative predictor retention. This guideline is theoretically supported by the sure screening property \citep{fan2008sure}, which suggests that SIS is more likely to capture true signals as the reduced dimension increases. Finally, three R packages are developed to optimize high-dimensional mediation models, with one package connecting R to the Intel oneAPI Math Kernel Library, significantly improving computational efficiency.

This article is structured to guide the reader through a comprehensive exploration of our innovative approaches to addressing internalizing psychopathology (IP). It begins with a motivating example in Section 2, which sets the stage for understanding the complexities involved. In Section 3, we compare conventional regularization methods with our novel techniques, providing additional justifications for our approach. Section 4 presents extensive simulation results that demonstrate the effectiveness of our proposed method, while Section 5 illustrates its application within the context of the IP study. Finally, the concluding section summarizes the key findings and suggests potential directions for future research, ensuring a well-rounded discussion of the topic.

\section{Motivational Example}
\label{ch:IP}

This work draws inspiration from a study conducted by the University of Illinois at Chicago, which explored the effects of various treatments on individuals experiencing internalizing psychopathology (IP). The IP spectrum includes major depressive disorder (MDD), generalized anxiety disorder (GAD), post-traumatic stress disorder (PTSD), and other related conditions. In this study, participants were divided into two groups: the untreated IP group and the healthy control group (HC). Individuals with IP were randomly assigned to receive 12 weeks of treatment, either through Selective Serotonin Reuptake Inhibitors (SSRIs) or Cognitive Behavioral Therapy (CBT). To assess the impact of these treatments, psychiatric evaluations were conducted both before and after the intervention, employing the Depression Anxiety Stress Scales (DASS) \citep{brown1997psychometric} to quantify changes in symptoms.

The study employed FC measures to identify disrupted brain connectivity in individuals with IP compared to a healthy control group. FC, which reflects the dynamic interactions among various brain regions during different mental processes, was assessed using resting-state functional Magnetic Resonance Imaging (rs-fMRI) data. The brain network was divided into 105 regions of interest (ROIs) based on the CONN atlas \citep{whitfield2012conn}, with detailed parcellation provided in the supporting information, resulting in a total of 5,460 distinct FC links. Additional information regarding participant demographics, selection criteria, data processing, and imaging acquisition can be found in previously published articles \citep{thomas2022network} related to this research.

Antidepressants provide rapid treatment effects over short durations, prompting our analysis to focus on identifying potential biomarkers for neuromodulation that could expedite recovery in patients undergoing CBT or those with severe symptoms of IP. Our study examines 28 participants who received antidepressants (the IP-SSRI group) and completed resting-state functional Magnetic Resonance Imaging (rs-fMRI) scans both before and after treatment. This group is compared to 27 participants in the HC group. By developing a mediation model, we aim to uncover neuroimaging biomarkers, particularly FC links, that mediate the relationship between treatment effects and improvements in neurobehavioral outcomes.

\section{Method}
The high-dimensional mediation model introduced in this study is pivotal for understanding how mediators influence the connection between treatment and outcome. By identifying mediators that demonstrate nonzero indirect effects (IE), this model facilitates the discovery of potential biomarkers. The framework for mediation is clearly defined, and an innovative optimization method is presented, which combines feature selection with penalized estimation to enhance the model's effectiveness.
\subsection{Model Definition}
We develop a mediation model utilizing the linear structural equation modeling technique (LSEM). The model encompasses $n$ subjects, each assumed to be independent and identically distributed. Within this framework, $X$ denotes the treatment exposure, represented as a binary variable (with $X=0$ for the HC group and $X=1$ for the IP-SSRI group), while $Y$ serves as the continuous outcome variable, reflecting changes in DASS scores by calculating the difference between post-treatment and pre-treatment scores. The mediator vector, represented as $M = (M_1, M_2, \ldots, M_p)$, captures the changes in FC that occur following the treatment.

\begin{equation} \label{eq:mediation_ch4}
 \left\{\begin{array}{@{}l@{}}
 M_{i} = X\alpha_i + \varepsilon_i, \quad i = 1,2,\ldots,p \\
 Y = X\gamma + \sum_{i=1}^{p} M_{i}\beta_i + \zeta
 \end{array}\right.\,.
\end{equation}

Errors associated with each mediator, $\varepsilon_i$, are assumed to follow a multivariate normal distribution with a mean of zero and a covariance matrix $\Sigma$. The error of the outcome model, $\zeta$, is assumed to follow a normal distribution with a mean of zero and a variance of $\sigma^2$. Furthermore, the errors associated with the mediators and the outcome variable are treated as independent.

Our mediation model assumes a parallel design \citep{jones2015health}, forgoing the incorporation of potential sequential relationships among mediators. This simplification is necessitated by insufficient empirical evidence in the extant literature supporting specific temporal orderings. Moreover, as demonstrated by \cite{zhao2022pathway}, the parallel mediation model can be seen as a simplified version of a sequential mediation model under certain conditions. As a result, the interpretation of individual mediation effects remains consistent, regardless of the underlying  structure or sequential dependencies among mediators.

\subsection{Mediation Effect}
The parameter vector $\bm{\alpha} = (\alpha_1, \alpha_2, \ldots, \alpha_p)$, of dimension $1 \times p$, characterizes the relationship between the treatment exposure $X$ and the mediator vector $M$, encapsulating the treatment effect on the variations in the FC mediators. Additionally, $\bm{\beta} = (\beta_1, \beta_2, \ldots, \beta_p)$, configured as a $p \times 1$ vector, describes the relationship between the mediator $M$ and the outcome variable $Y$, quantifying the influence of FC mediators on the improvement in neurobehavioral outcomes.

Our mediation model, grounded in the above framework for potential outcomes \citep{rubin1974estimating}, facilitates the examination of multiple mediation pathways while maintaining a clear delineation between DE and IE. This framework decomposes TE on an outcome into path-specific effects \citep{avin2005identifiability,daniel2015causal} operating through distinct mediators. The TE comprises two primary components: the DE and the TIE. The DE of treatment exposure $X$ on outcome $Y$, without the influence of mediators, is quantified by the parameter $\gamma$. The TIE, also termed the mediation effect, representing the influence of treatment on the outcome through mediators, is computed using the product-coefficient method as $ \sum_{i=1}^{p} \alpha_i \times \beta_i$. Within this formulation, the IE through the $i$th mediator, denoted as $IE_i$, is expressed as $\alpha_i \times \beta_i$. This approach allows for the decomposition of the TIE into individual path-specific effects, enabling a nuanced analysis of complex mediation structures and providing insights into the relative contributions of each mediator to the TIE.

\subsection{ Estimation of Parameters} \label{subsec:PM}
This  research aims to improve the detection of  mediators with nonzero IE. We adopt a simpler approach to stream the estimation process by assuming a unit variance for the error distribution rather than employing a more intricate covariance matrix. This simplification will not compromise the consistency of the least-square estimators, provided that all variables are standardized to a unit scale \citep{white1980heteroskedasticity}. Accordingly, the log-likelihood $l(\bm{\alpha}, \bm{\beta}, \gamma )$, is specified as follows:
\begin{equation}
\begin{split}
l(\bm{\alpha}, \bm{\beta}, \gamma )=tr \{(\bm{M-X\alpha})^{t}(\bm{M-X\alpha})\}+
(\bm{Y-X}\gamma-\bm{M\beta})^{t}(\bm{Y-X}\gamma-\bm{M\beta}).
\end{split}
\end{equation}

\subsubsection{Commonly used  Methods}
Two traditional regularization methods, LASSO and Pathway LASSO, are utilized to estimate IE for this the high-dimensional model. When applied to our model, the LASSO method can be viewed in terms of the following equation:
\begin{equation}
\underset{\bm{\alpha}, \bm{\beta}, \gamma}{\mathrm{argmin}} \{ \frac{1}{2}l(\bm{\alpha}, \bm{\beta}, \gamma ) + P_{1}(\bm{\alpha}) + P_{2}(\bm{\beta})\}.
\end{equation}

The $L_1$-norm penalty functions are defined as $P_{1}(\bm{\alpha}) = \lambda_{1\alpha} \sum_{i=1}^{p} \lvert \alpha_i\rvert$ and $P_{2}(\bm{\beta}) = \lambda_{1\beta} \sum_{i=1}^{p} \lvert \beta_i \rvert$. These functions impose penalties on the model parameters $\bm{\alpha}$ and $\bm{\beta}$ to enhance the accuracy of parameter selection.

The Pathway LASSO method can be viewed in terms of the following equation, which is aimed at stabilizing estimates and minimizing estimation bias:
\begin{equation}
 \underset{\bm{\alpha}, \bm{\beta}, \gamma}{\mathrm{argmin}} \{ \frac{1}{2}l(\bm{\alpha}, \bm{\beta}, \gamma ) + P_{1}(\bm{\alpha}, \bm{\beta}) + P_{2}(\bm{\alpha, \beta}) + P_{3}(\gamma)\}.
\end{equation}

This method, drawing from the principles of the EN, introduces a penalty $P_{1}(\bm{\alpha, \beta}) = \kappa \sum_{i=1}^{p} [\lvert \alpha_i \beta_i\rvert + \nu(\alpha_i^2+ \beta_i^2)]$ for the identification of path-specific effect, and this penalty remains convex when $\nu$ is greater than or equal to 0.5.  An additional $L_1$-norm penalty $P_{2}(\bm{\alpha, \beta}) = \omega \sum_{i=1}^{p} (\lvert \alpha_i\rvert + \lvert \beta_i \rvert)$ is included to further shrink the individual parameters $\alpha_i$ and $\beta_i$, thereby improving the selection accuracy. Moreover, to address the overestimation of the DE ($\gamma$), an $L_1$-norm penalty for $P_{3}(\gamma) = \lambda_{\gamma} \lvert \gamma\rvert$ is also proposed in this method.

\subsubsection{The Proposed Method}
We propose two tuning parameter strategies that relax the $L_1$-norm constraints to address challenges in traditional methods. The first strategy introduces an $L_1$-norm penalty $\lambda_{\gamma} \lvert \gamma\rvert $ for the DE and establishes a minimum threshold for $\lambda_\gamma$ (specified as $\lambda_\gamma \geq c$) to address the issue of overestimation. This adjustment is particularly effective in reducing DE overestimation by compressing its estimated value towards zero, leading to a lower bias and variance of the estimated IE.

The second strategy aims to further enhance the TPR of IE by critically evaluating parameter estimation in IE ($\alpha_i\beta_i$) and improving the detection of mediators with nonzero impacts. This is achieved by reducing the magnitude of the tuning parameter, $\lambda_{1\beta}$, which is linked to the $L_1$-norm penalty for the parameter vector $\bm{\beta}$, under the condition that $\lambda_{1\alpha} > \lambda_{1\beta}$. The justification for this strategy is detailed subsequently, focusing on the influence of the tuning parameters $\lambda_{1\alpha}$ and $\lambda_{1\beta}$ on parameter estimation. 

\begin{itemize}
 \item The parameter vector $\bm{\alpha}$ consistently exhibits high TPR due to the unique structure of the mediation model. Each $\alpha_i$ is independently estimated in its mediator model, see Equation \ref{eq:mediation_ch4}, making the detection nonzero values of $\alpha_i$ stable and less affected by the magnitude of the penalty $\lambda_{1\alpha}$, even for $\alpha_i$ values with smaller strengths.
 
 \item The parameter $\beta_i$ in the outcome model (Equation \ref{eq:mediation_ch4}) is more susceptible to the estimation bias of $\alpha_i$ \citep{jones2015health}, the magnitude of tuning parameter $\lambda_{1\beta}$, and the outcome noise ($\zeta$), leading to higher MSE of the parameter $\bm{\beta}$ than $\bm{\alpha}$. Shrinking the size of $\lambda_{1\beta}$ increases the likelihood of detecting nonzero $\beta_i$ values, thus improving the identification of mediators with nonzero effects.
\end{itemize}

\subsection{ Evaluation of Models}
Tuning Parameters: We selected seven values for the tuning parameters $\lambda_{1\alpha}$ and $\lambda_{1\beta}$: $0.001$, $0.01$, $0.1$, $1$, $2$, $5$, and $10$. This resulted in 49 combinations of penalty parameters ($\lambda_{1\alpha},\lambda_{1\beta}$) being examined. Additionally, we varied $\lambda_\gamma$ across a spectrum of 72 values, ranging from $0$ to $100$. Hence, we explored 3528 parameter combinations for ($\lambda_{1\alpha}, \lambda_{1\beta}, \lambda_\gamma$), allowing for thoroughly exploring model behaviors under diverse settings.

Model Selection: To identify the optimal model, we utilized the Bayesian information criterion (BIC). The BIC is calculated using the formula $BIC = qln(n) - 2 ln(L)$, where $q$ denotes the total number of nonzero estimated parameters and $L$ is the estimated maximum likelihood. We opted for the BIC instead of cross-validation (CV) for the following reasons. First, when a large penalty is applied, all parameter estimates may be reduced to zero. Consequently, the average predictive error would be solely determined by noise, potentially resulting in the lowest predictive error. This situation could lead to the incorrect identification of the optimal model. Additionally, CV is more computationally intensive. Therefore, choosing BIC is not only theoretically and statistically justified, but it  is also practically meaningful.

Performance Metrics:  We evaluated the performance of the estimated IE   using the True Positive Rate (TPR) and True Negative Rate (TNR) to assess sensitivity and specificity, respectively. Mean Squared Error (MSE) and Relative Bias (RB) measured parameter estimation accuracy for $\hat{\bm{\alpha}}$, $\hat{\bm{\beta}}$, and $\hat{\alpha_i}\hat{\beta_i}$. Overall performance was assessed using the F1 score (harmonic mean of TPR and True Discovery Rate) and Youden's J statistic ($J = TPR + TNR - 1$), which combines sensitivity and specificity.

\subsection{Screening Procedure for Ultra-high Dimensional Mediators}
To enhance the accuracy of parameter estimation and reduce computational load, we adopt the SIS approach, which preserves essential information during the dimensional reduction process. In mediation analysis, SIS selects potential mediators based on their marginal correlations with the outcome. The reduced dimension is typically set at $d=k[n/log(n)]$. We hypothesize that the choice of $k$ will be influenced by the ratio of the number of mediators to the sample size. We conduct a simulation study to identify the optimal value of $k$, particularly in cases where the number of mediators changes while the sample size remains fixed.

\subsection{An Algorithm for Optimization and Development of R packages}
To estimate parameters, we employed the Alternating Direction Method of Multipliers (ADMM) \citep{boyd2011distributed} for the optimization problem. We have formulated an R package for the context of a high dimensional mediation model, designated as \texttt{HDMAADMM}, which implements ADMM with an SIS option for regularization. To expedite the large-scale computational tasks, we created two R packages, \texttt{oneMKL} and \texttt{oneMKL.MatrixCal}. These are designed to work seamlessly with the Intel oneAPI Math Kernel Library, offering the functions of matrix operations and parallel computing in R on both Linux and Windows platforms. 

\section{Simulation Study}
This section presents two simulation studies evaluating the effectiveness of our proposed tuning parameter strategies.

\subsection{ Designs for Simulations}
We simulated 200 independent and identically distributed samples. Each sample contained $n = 50$ subjects with a varying number of mediators $p$ = $30$, $50$, $100$, $150$, and $200$. The treatment exposure $X$ was created using a Bernoulli distribution with a probability of 0.5. Given that our main purpose was to estimate the IE, we set the true value for the DE parameter $\gamma$ to a relatively small value at 2. The parameters associated with IE is $\alpha_i\beta_i$ organized as follows: 

\begin{itemize}
 \item The vector $\bm{\alpha} = (\alpha_1, \alpha_2, \ldots, \alpha_p)$, representing the effect of treatment on the mediator, had its first 20\% of values as nonzeros, and the remaining 80\% were set at zero.
 
 \item The vector $\bm{\beta} = (\beta_1, \beta_2, \ldots, \beta_p)$, indicating the effect of the mediator on the outcome, was arranged differently. The first 10\% of parameters were given nonzero values, followed by 10\%, which were set to zero. The next 10\% were given nonzero values, and the remaining 70\% were set at zero.

 \item For nonzero values of parameters $\alpha_i$ and $\beta_i$, we categorized them into large signals (constituting 5\%, with a mean of 6 and a standard deviation of 0.1) and small-signals (constituting 5\%, the mean being 4 and the standard deviation of 0.1). Both types of signals adhered to normal distributions. 
\end{itemize}

Each mediator $M_i$ (for $i = 1, 2, \ldots,p$) was derived using Equation \ref{eq:mediation_ch4}, where the error distribution for the mediators adhered to a multivariate normal distribution with a mean vector of 0. For simplicity, the covariance matrix of mediators was defined as a compound symmetry matrix, with diagonal elements assigned to a value of 1 and off-diagonal elements assigned to another value of 0.1. This specification was informed by detecting very weak correlations among the FC mediators observed in the IP study. Given that the IE for the $i$th mediator is defined as $IE_i = \alpha_i \beta_i$, we categorized mediators into four distinct types based on the nonzero status of their component parameters. Type 1 mediators (10\%) exhibit nonzero IE ($\alpha_i \neq 0$, $\beta_i \neq 0$), creating an $X-M-Y$ pathway. The remaining 90\% are zero-effect mediators: Type 2 ($\alpha_i = 0$, $\beta_i \neq 0$) creates an $M-Y$ pathway, Type 3 ($\alpha_i \neq 0$, $\beta_i = 0$) forms an $X-M$ pathway, and Type 4 ($\alpha_i = 0$, $\beta_i = 0$) shows no effect.

The outcome variable $Y$ is generated using Equation \ref{eq:mediation_ch4}, where its error distribution follows a normal distribution with a mean of zero and a standard deviation of 0.1. To avoid estimation problems, particularly for mediators with small nonzero signals, we ensured that the standard deviation of the error for the outcome and the standard deviation of the nonzero parameter $\alpha_i$ and $\beta_i$ were on the same scale. This approach helps get accurate estimates.

\subsection{Simulation Results}
\subsubsection{Overestimation of DE}

We initiate our analysis by examining the effectiveness of our first tuning parameter strategy in mitigating the overestimation of the DE. In the simulation study, we compare the LASSO method that utilizes three distinct tuning parameter strategies, specifically, (1) $TR_L$: the traditional LASSO method, without applying any penalty to the DE (where the tuning parameter $\lambda_\gamma = 0$), (2) $MD_L$: the LASSO method with a modified $L_1$-norm penalty $ \lambda_{\gamma} \lvert \gamma\rvert$ (where $\lambda_\gamma > 0$) for the DE, and (3) $SMD_L$: the LASSO method with a strictly modified $L_1$-norm penalty $ \lambda_{\gamma} \lvert \gamma\rvert$ (where $\lambda_\gamma \geq c$) for the DE. This comparison examines how various tuning parameter strategies impact the accuracy of estimating parameters. Additionally, we report the findings by altering the number of mediators ($p$) across various levels. The optimal model is determined by BIC.

An example illustrates the overestimation of the DE when the number of mediators is 200. In this case, we assume the true value of TE to be 521.736, which consists of a DE of 2 and an IE of 519.736. Using the $TR_L$ strategy, the average estimation of DE tends to be greatly overestimated, reaching 505.382. IE is substantially underestimated at 12.998, leading to a poor TPR of IE (18.6\%). 

The $MD_L$ strategy significantly reduces the overestimation of the DE to 312.117. Even though this estimate is substantially lower than the overestimation observed by the $TR_L$ strategy, it is still undesirable as the true value of DE is $2$. The estimated IE increases from 12.998 to 194.406. The MSE of IE decreases from 82.094 to 68.408. The RB is also reduced, moving from -0.975 to -0.626, and the TPR of IE increases from 18.6\% to 46.4\%. 

Among the three tuning parameter strategies evaluated, the overall performance of the $SMD_L$ strategy demonstrates superior overall performance compared to the alternative approaches. Using grid search, when threshold $c$ is set to 0.3, the $SMD_L$ strategy reduces the estimated DE to 1.688, closely approximating the true DE value of 2. As a result, the $SMD_L$ strategy elevates the TPR to 81.1\%, reduces the MSE to 64.690, and lowers the RB to -0.059. Thus, adjusting the range of the tuning parameter $\lambda_\gamma$ beyond 1 is crucial for achieving the desired level of TPR of IE in high-dimensional mediation models.

\begin{figure}[h]
\centering
\includegraphics[width=1.0\textwidth]{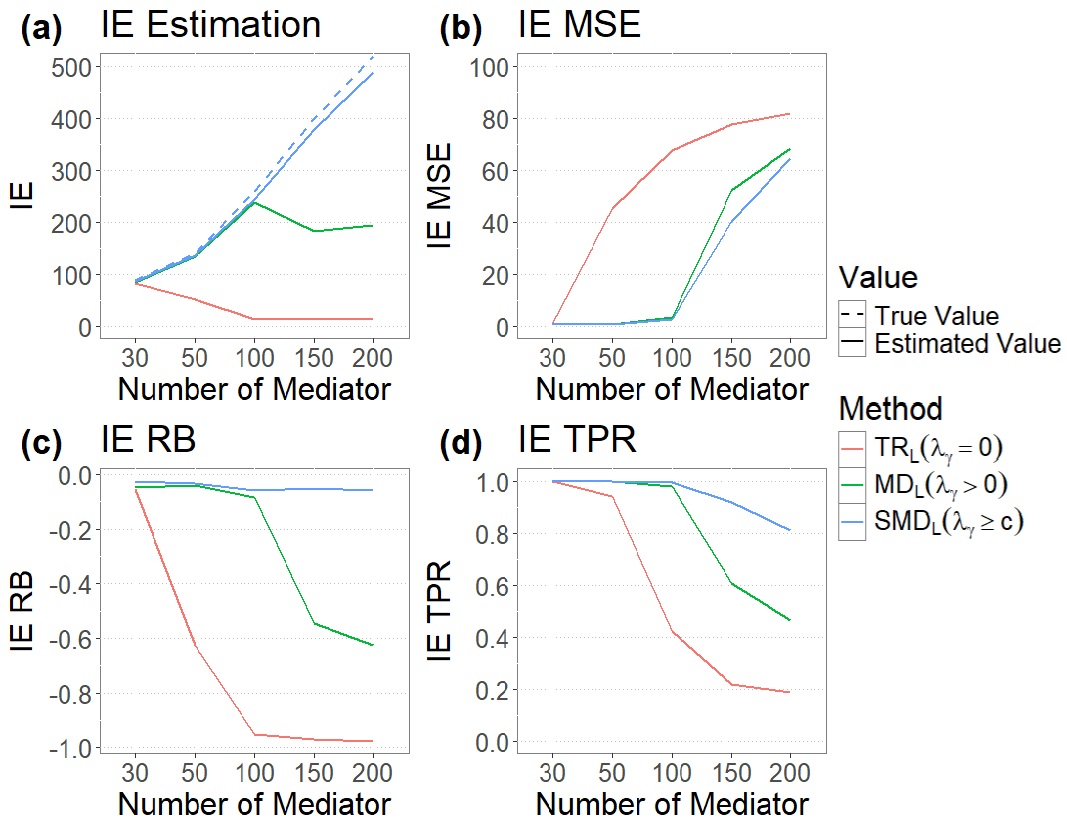}
\caption{ Simulated Results by LASSO - Overestimation of DE}
\label{fig:LASSO_Overestimate}
\end{figure}

Figure \ref{fig:LASSO_Overestimate} presents simulation results with varying numbers of mediators. In a low-dimensional model setting (with $n = 50$ and $p = 30$), the $TR_L$ strategy accurately estimates the IE. However, as the number of mediators increases, the DE obtained by the $TR_L$ strategy is substantially overestimated. In high-dimensional settings (with $n = 50$ and $p \ge 50$), the impact of the tuning parameter strategy on the estimation is consistent with the previous example ($p = 200$). The $TR_L$ strategy overestimates DE substantially, resulting in an estimation of IE close to zero, a larger estimation of MSE and RB, and a smaller estimation of TPR and TNR. Conversely, the $MD_L$ and $SMD_L$ strategies successfully mitigate this overestimation, enhancing the accuracy of estimation and biomarker detection. Notably, the $SMD_L$ strategy is particularly beneficial in estimating IE with desirable MSE, RB, TPR, and TNR.

The abovementioned findings assumed a small true value of $2$ for DE. To further examine the $SMD_L$ strategy, we conducted another simulation study with a sample size of 50 and 200 mediators, mainly focusing on situations with larger DE values. In this simulation, the true value of IE was fixed at 320, while the true values of DE were varied across a spectrum of $36, 80, 137, 213, 320, 480, 747, 1280, 2880$, and $6080$. This spectrum represents a transition in the mediation proportion (i.e., the ratio of IE to TE) from 90\% to as low as 5\%. 

Simulation results for the $MD_L$ and $SMD_L$ strategies showed an overestimation bias in IE, intensifying as the true value of DE increased. The underlying reason for this phenomenon is that the TE is a sum of DE and TIE; when the estimation of DE is constrained, the estimation of TIE correspondingly increases, leading to a rise in both the MSE and RB of IE. Moreover, a threshold $c$ in the $SMD_L$ strategy was ascertained by identifying the minimal MSE of DE derived from the grid research outcomes. This strategy facilitates an accurate estimation of DE and enhances the TPR of IE.

It is noteworthy that the TPR of the IE under the $SMD_L$ strategy  ranges between 69\% and 81\%, which is significantly larger compared to the $MD_L$ strategy (ranging from 46\% to 60\%) and the $TR_L$ strategy (ranging from 16\% to 27\%). The TNR of IE for the three strategies exhibits similar values. To conclude, when the research objective is to enhance biomarker detection rather than providing a precise estimate of IE, the $SMD_L$ strategy remains an effective strategy for addressing the problem of overestimation of DE.

\subsubsection{Improving the Detection Rate  of Nonzero-Effects of  Mediators}

Another simulation study examined whether the second tuning parameter strategy further improves model performance. We relaxed the assumption of the $L_1$-norm equal-sized tuning parameter constraint (where $\lambda_{1\alpha} = \lambda_{1\beta}$) for parameters regarding IE (i.e., $\bm{\alpha}$ and $\bm{\beta}$) and observed whether this relaxation contributes to the improvment of model performance. 

Table \ref{tab:Example} illustrates the impact of this relaxation of the constraint using an example with 20 nonzero mediators, with the  estimation of parameters obtained via the $SMD_L$ strategy.  When penalty sizes are equal, such as $\lambda_{1\alpha} = 1$ and $\lambda_{1\beta} = 1$, we observe that only 12 out of 20 nonzero mediators are detected, yielding a TPR of the IE of $60\%$. Reducing $\lambda_{1\alpha}$ to $0.1$ while keeping $\lambda_{1\beta}$ at 1 (where $\lambda_{1\alpha} < \lambda_{1\beta}$) does not improve the TPR of IE, as it does not enhance the TPR of $\bm{\beta}$. However, lowering $\lambda_{1\beta}$ to $0.01$ while keeping $\lambda_{1\alpha}$ at 1 leads to the detection of 10 additional nonzero $\beta_i$, raising the TPR of IE to $95\%$. 

\begin{table}[h]
\caption{ Performance of $SMD_L$ under various combinations of $\lambda_{1\alpha}$ and $\lambda_{1\beta}$}
\label{tab:Example}
\centering
\begin{adjustbox}{scale=0.80}
\begin{tabular}{@{}cc|rrr|rrr|rrr|rrrrr@{}}
\hline
\multicolumn{5}{c}{ \multirow{2}{*}{Tuning Parameter Strategy}}&\multicolumn{3}{c}{$\lambda_{1\alpha} = \lambda_{1\beta}$} & \multicolumn{3}{c}{$\lambda_{1\alpha} < \lambda_{1\beta}$} & \multicolumn{3}{c}{$\lambda_{1\alpha} > \lambda_{1\beta}$} \\ \cline{6-14}
\multicolumn{5}{c}{}&\multicolumn{3}{c}{$(\lambda_{1\alpha} = 1, \lambda_{1\beta} = 1)$} & \multicolumn{3}{c}{$(\lambda_{1\alpha} = 0.1, \lambda_{1\beta} = 1)$} & \multicolumn{3}{c}{$(\lambda_{1\alpha} = 1, \lambda_{1\beta} = 0.01)$} \\ \cline{1-14}
\multirow{2}{*}{Signal}& \multirow{2}{*}{Mediator}&\multicolumn{3}{c}{True Value} & \multicolumn{9}{c}{Estimation}\\  \cline{3-14}
 & & \multicolumn{1}{c}{$\alpha$} & \multicolumn{1}{c}{$\beta$}& \multicolumn{1}{c}{$IE$} & \multicolumn{1}{c}{$\alpha$} & \multicolumn{1}{c}{$\beta$}& \multicolumn{1}{c}{$IE$} &\multicolumn{1}{c}{$\alpha$} & \multicolumn{1}{c}{$\beta$}& \multicolumn{1}{c}{$IE$} &\multicolumn{1}{c}{$\alpha$} & \multicolumn{1}{c}{$\beta$}& \multicolumn{1}{c}{$IE$} \\ \hline

 \multirow{10}{*}{Large} &1 &6.073 & 5.936 & 36.049 & 6.115 & 4.901 & 29.971 & 6.236 & 4.901 & 30.562 & 6.115 & \bf{4.029} & 24.634\\
 & 2 & 6.141 & 6.016 & 36.944 & 5.780 & 1.696 & 9.804 & 5.894 & 1.696 & 9.997 & 5.780 & \bf{3.583} & 20.711 \\
 & 3 & 5.912 & 5.823 & 34.426 & 5.515 & 0.000 & 0.000 & 5.626 & 0.000 & 0.000 & 5.515 & \bf{0.632} & 3.486 \\
 & 4 & 5.899 & 5.758 & 33.968 & 5.869 & 0.000 & 0.000 & 5.987 & 0.000 & 0.000 & 5.869 & \bf{1.099} & 6.452 \\
 & 5 & 6.160 & 5.827 & 35.898 & 5.606 & 0.000 & 0.000 & 5.718 & 0.000 & 0.000 & 5.606 & \bf{3.539} & 19.841 \\
 & 6 & 6.168 & 5.853 & 36.101 & 6.065 & 9.064 & 54.975 & 6.185 & 9.064 & 56.057 & 6.065 & \bf{5.618} & 34.074\\ 
 & 7 & 5.929 & 5.913 & 35.054 & 5.992 & 10.494 & 62.881 & 6.111 & 10.494 & 64.131 & 5.992 & \bf{4.430} & 26.547 \\
 & 8 & 6.091 & 5.989 & 36.482 & 5.667 & 12.946 & 73.361 & 5.779 & 12.946 & 74.811 & 5.667 & \bf{13.882} & 78.666 \\
 & 9 & 6.084 & 5.977 & 36.364 & 6.206 & 9.021 & 55.989 & 6.329 & 9.021 & 57.092 & 6.206 & \bf{7.999} & 49.644 \\
 & 10 &5.850 & 5.922 & 34.644 & 6.024 & 0.000 & 0.000 & 6.142 & 0.000 & 0.000 & 6.024 &\bf{0.456} & 2.746 \\\cline{1-14}
 \multirow{10}{*}{Small} & 11 & 4.009 & 3.968 & 15.907 & 4.026 & 0.144 & 0.580 & 4.111 & 0.144 & 0.593 & 4.026 & \bf{0.753} & 3.032 \\ 
 & 12 &4.075 & 4.180 & 17.034 & 4.216 & 0.000 & 0.000 & 4.301 & 0.000 & 0.000 & 4.216 & \bf{2.936} & 12.379\\ 
 & 13 & 4.188 & 4.061 & 17.007 & 4.043 & 2.021 & 8.171 & 4.125 & 2.021 & 8.336 & 4.043 & \bf{3.576} & 14.457 \\
 & 14 & 4.101 & 3.966 & 16.267 & 3.883 & 3.989 & 15.488 & 3.966 & 3.989 & 15.818 & 3.883 & \bf{2.553} & 9.914 \\
 & 15 & 3.934 & 3.892 & 15.315 & 3.894 & 5.549 & 21.607 & 3.975 & 5.549 & 22.060 & 3.894 & \bf{2.602} & 10.131 \\
 & 16 & 3.813 & 4.146 & 15.807 & 4.012 & 6.667 & 26.745 & 4.095 & 6.667 & 27.305 & 4.012 & \bf{5.908} & 23.701 \\
 & 17 &4.123 & 3.949 & 16.280 & 3.678 & 0.000 & 0.000 & 3.756 & 0.000 & 0.000 & 3.678 & \bf{0.000} & 0.000 \\
 & 18& 4.023 & 3.935 & 15.832 & 4.313 & 0.000 & 0.000 & 4.403 & 0.000 & 0.000 & 4.313 & \bf{1.898} & 8.184 \\
 & 19 &4.130 & 4.171 & 17.224 & 4.071 & 4.596 & 18.709 & 4.159 & 4.596 & 19.113 & 4.071 & \bf{2.473} & 10.067 \\
 & 20 &3.740 & 3.953 & 14.783 & 4.022 & 0.000 & 0.000 & 4.107 & 0.000 & 0.000 & 4.022 & \bf{1.003} & 4.035\\\hline
  \multicolumn{2}{l}{\multirow{3}{*}{IE Performance}} &\multicolumn{3}{l}{MSE} &0.060 & 5.280 & 71.666 & 0.070 & 5.280 & 73.887 & 0.060 & 3.470 & \bf{47.907} \\ 
 \multicolumn{2}{c}{} &\multicolumn{3}{l}{TNR}&0.138 &0.969 & 0.972 & 0.013 & 0.969 & 0.972 & 0.175 & 0.819 & \bf{0.806} \\
\multicolumn{2}{c}{} &\multicolumn{3}{l}{TPR} &1.000 & 0.300 & 0.600 & 1.000 & 0.300 & 0.600 & 1.000 & 0.800 & \bf{0.950} \\\hline

\end{tabular}
\end{adjustbox}
\end{table}

Given that the sum of DE and TIE is a fixed number (TE), the accuracy of estimating DE will inherently influence the estimation outcomes of IE. Consequently, the first tuning parameter strategy ($\lambda_{\gamma} \lvert \gamma\rvert$ where $\lambda_\gamma \geq c$), designed to determine the estimation of DE, needs to be taken into account when evaluating the second strategy. Thus, a total of 4 methods were compared in this simulation study using the Pathway LASSO method. 

Four variants of the Pathway LASSO method were implemented: $SMD.S_P$ (where $\lambda_\gamma \geq c$, $\lambda_{1\alpha} > \lambda_{1\beta}$), $SMD.E_P$ ($\lambda_\gamma \geq c$, $\lambda_{1\alpha} = \lambda_{1\beta}$), $MD.S_P$ ($\lambda_\gamma > 0$, $\lambda_{1\alpha} > \lambda_{1\beta}$), and $MD.E_P$ ($\lambda_\gamma > 0$, $\lambda_{1\alpha} = \lambda_{1\beta}$), with the traditional Pathway LASSO equivalent to $MD.E_P$. For a consistent comparison, the tuning parameters $\kappa$ and $\nu$ in the path-specific penalty $P_{1}(\bm{\alpha, \beta}) = \kappa \sum_{i=1}^{p} [\lvert \alpha_i \beta_i\rvert + \nu(\alpha_i^2+ \beta_i^2)]$ were fixed across all methods. The parameter $\kappa$, crucial in determining the magnitude of estimated IE, was set to 0.01 to optimize results, leading to higher TPR and lower MSE of the IE, as well as reduced BIC of the model. The parameter $\nu$ was fixed at 2, adhering to recommendations from the traditional Pathway LASSO.

To evaluate the  performance of the proposed model  for the aferomentioned  four methods, we refrained from identifying an optimal model among the 3,528 tuning parameter combinations ($\lambda_\gamma, \lambda_{1\alpha}, \lambda_{1\beta}$). We noted the potential bias induced by the substantial variance in BIC across various penalty configurations ($\lambda_{1\alpha}, \lambda_{1\beta}$). Therefore, we employed a two-stage model selection process. The first stage involves selecting the optimal model based on the BIC for each sample, considering different penalty configurations ($\lambda_{1\alpha}, \lambda_{1\beta}$). The second stage aggregated the results by computing the average performance across the collection of the optimal model and penalty configurations within each method. This selection method aligned with the Minimax property principle, which ensured a more robust and reliable assessment of model performance.

\begin{figure}[h]\centering
\captionsetup{justification=centering}
\includegraphics[width=1.0\textwidth]{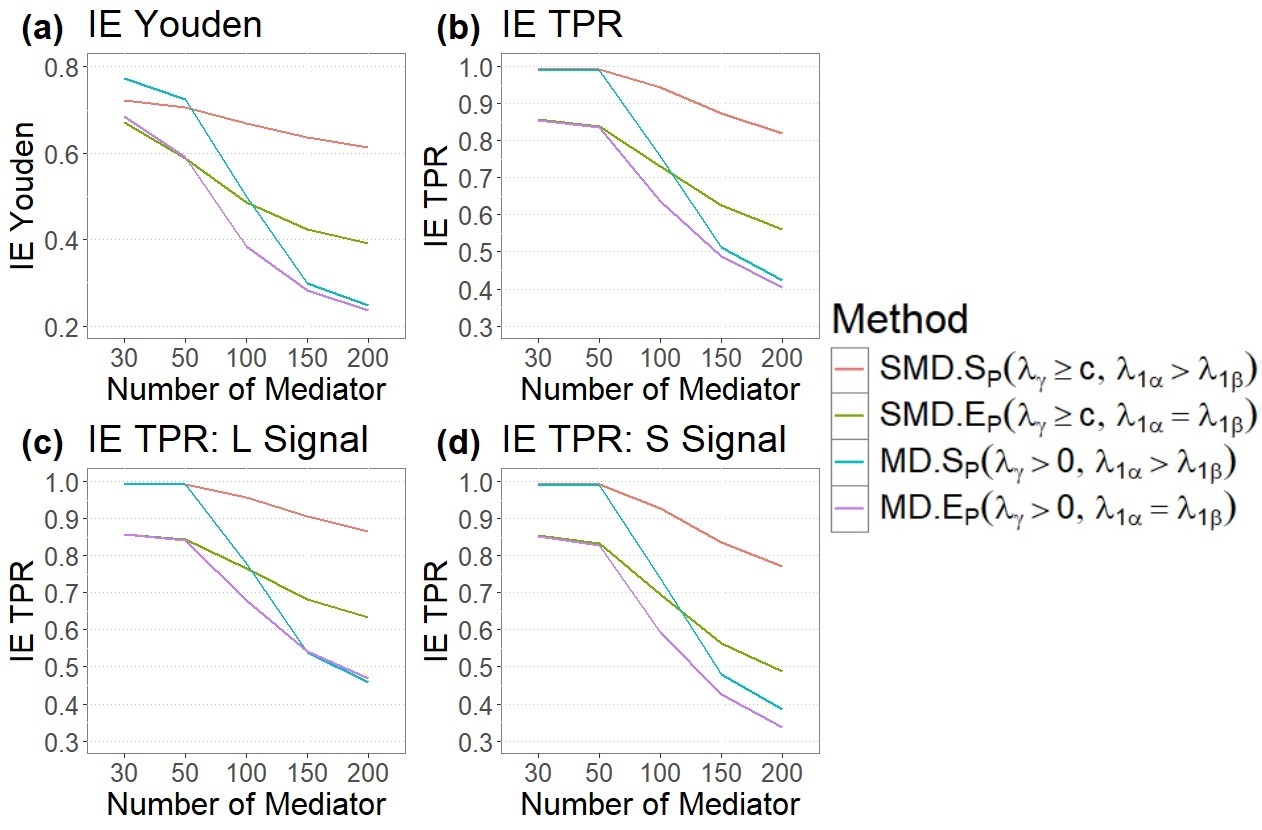}
\caption{A Comparison between SMD and MD Using Pathway LASSO for Various Combinations of $\lambda_{1\alpha}$ and $\lambda_{1\beta}$}
\label{fig:methodcomparison}
\end{figure}

Figure \ref{fig:methodcomparison} displays the simulation results with varying numbers of mediators. These results indicate that implementing our proposed tuning parameter strategies, $SMD.S_p$, exhibits superior  performance. Specifically, as the number of mediators increases, $SMD.S_p$ and $SMD.E_p$, employing our first tuning parameter strategy for DE (i.e., setting $\lambda_\gamma \geq c$) show better performance compared to those that do not  apply this strategy. The current finding underscores the effectiveness of our first strategy over the second strategy, which involves setting a constraint on the tuning parameter for IE (i.e., setting $\lambda_{1\alpha} > \lambda_{1\beta}$). 

We further evaluated the effectiveness of the second tuning parameter strategy in enhancing model performance. Based on the simulation results for a scenario involving 200 mediators, we can infer that the methods implemented with a smaller size of the tuning parameter $\lambda_{1\beta}$, (specifically, $MD.S_P, SMD.S_P$), demonstrate a better model performance. This improvement is evident showing smaller MSE compared to cases where the penalty size for $\lambda_{1\alpha}$ and $\lambda_{1\beta}$ is equal (namely, methods $MD.E_P$ and $SMD.E_P$). This finding suggests that our second tuning parameter strategy, which includes reduced $\lambda_{1\beta}$, tends to yield estimates (particularly for parameters $\bm{\beta}$) that are closer to the corresponding actual model parameters. It accurately identifies the mediators with nonzero-effects. Our proposed method, $SMD.S_P$, outperforms others across IE metrics, with the highest TPR (77.1\%), F1 score (0.504), Youden Index (0.614), and lowest MSE (45.240). For signal detection, $SMD.S_P$ excels with TPRs of 86.7\% for large signals and 77.1\% for small signals, surpassing other methods in identifying both magnitudes.

\subsubsection{Sure Independence Screening for Finding Optimal Scaling Factor}
In this section, we conduct a simulation study to test our hypothesis that the choice of the scaling factor $k$ in the SIS procedure is influenced by the ratio of the number of mediators ($p$) to the sample size ($n$), referred to as the $p/n$ ratio. For this simulation, we keep the sample size constant at 50 while changing the number of mediators to 200, 500, and 1000. This gives us $p/n= 4, 10, 20$. We calculated the reduced dimension as $d=k[n/log(n)]$, and adjusted the scaling factor $k$ across a range of values: $0, 1, 2, 3, 4, 5, 6, 7, 8, 9, 10, 15, 20, 30$, and $40$. 

\begin{figure}[h]\centering
\captionsetup{justification=centering,margin=2cm}
\includegraphics[width=0.95\textwidth]{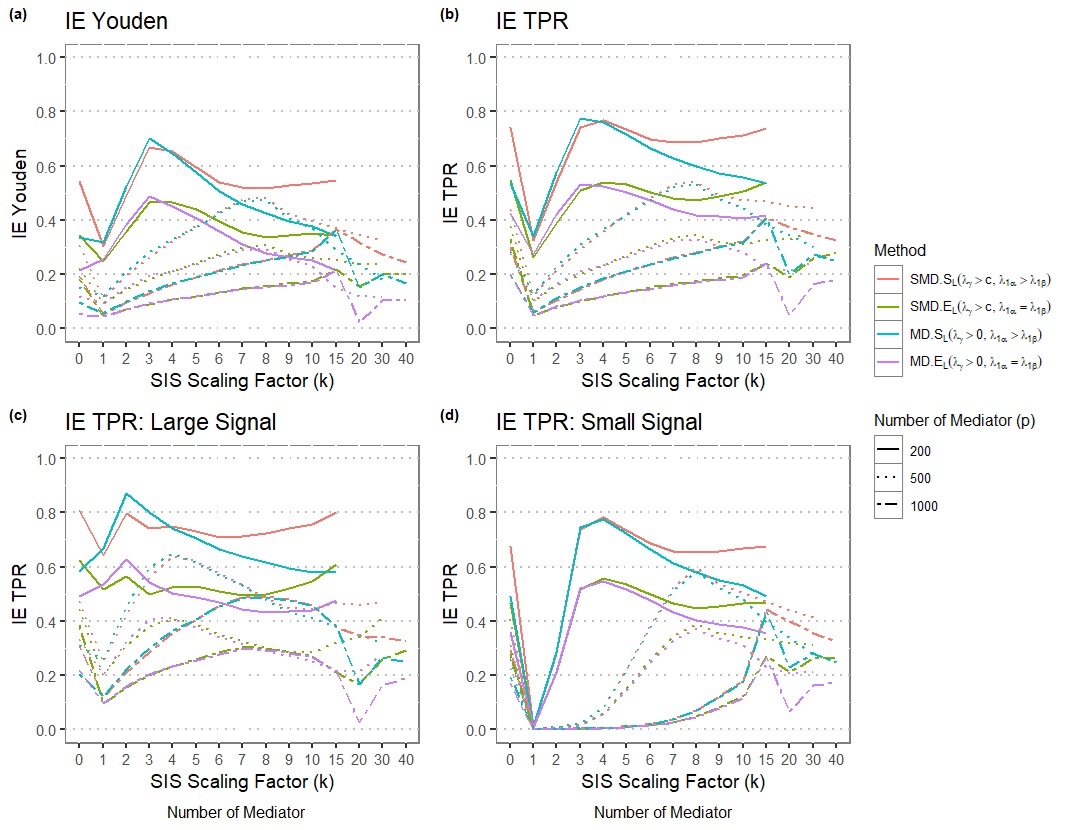}
\caption{Simulated Results for SIS}
\label{fig:SIS}
\end{figure}

Figure \ref{fig:SIS} presents simulated results for the varying scaling factor $k$. When the number of mediators was set at 200, yielding a $p/n$ ratio to 4, the largest TPR for identifying mediators with nonzero effects is attained at $k=3$ and $4$. This corresponds to reduced dimensions of 38 and 51, respectively. When the number of mediators is increased to 500, leading to a $p/n$ ratio of 10, the TPR peaks at $k = 8$, giving a reduced dimension of mediators at 102. In the case of 1000 mediators, where the $p/n$ ratio stands at 20, the best TPR occurs at $k = 15$ or $k = 20$, resulting in reduced dimensions of 192 or 256, respectively. These findings support our proposition that the $p/n$ ratio influences the optimal choice of $k$. The understanding of the motivational forces behind this is that when the reduced dimension $d$ is increased, the sure screening property of the SIS method is more likely to capture a greater number of true signals. Based on the findings from our simulations, with a fixed sample size of $n = 50$, it is advisable for researchers to select the $k$ value in accordance with the $p/n$ ratio.

\section{A Real World Application — IP Study}
We applied our proposed approach, denoted as $SMD.S_p$ (symbolizing the application of Pathway LASSO integrated with the two proposed tuning parameter strategies), to identify FC mediators with nonzero IE in the IP study. The primary objective is to elucidate the mediating role of FC between anti-depressant treatment and the alterations in neurobehavioral outcomes. To improve the TPR of IE and ensure computational feasibility, the SIS framework was implemented, utilizing a scaling factor of $k=15$. This adjustment significantly decreased the number of FC mediators we examined from 5460 to 206, aligning with the common practice of assuming 2-5\% signal sparsity. Through the screening procedure, the remaining FC revealed a much stronger association between the improvement in DASS scores and changes in FC within the IP group, taking into account the baseline FC levels of the HC group.

\begin{figure}[h]
\centering
\includegraphics[width=1.0\textwidth]{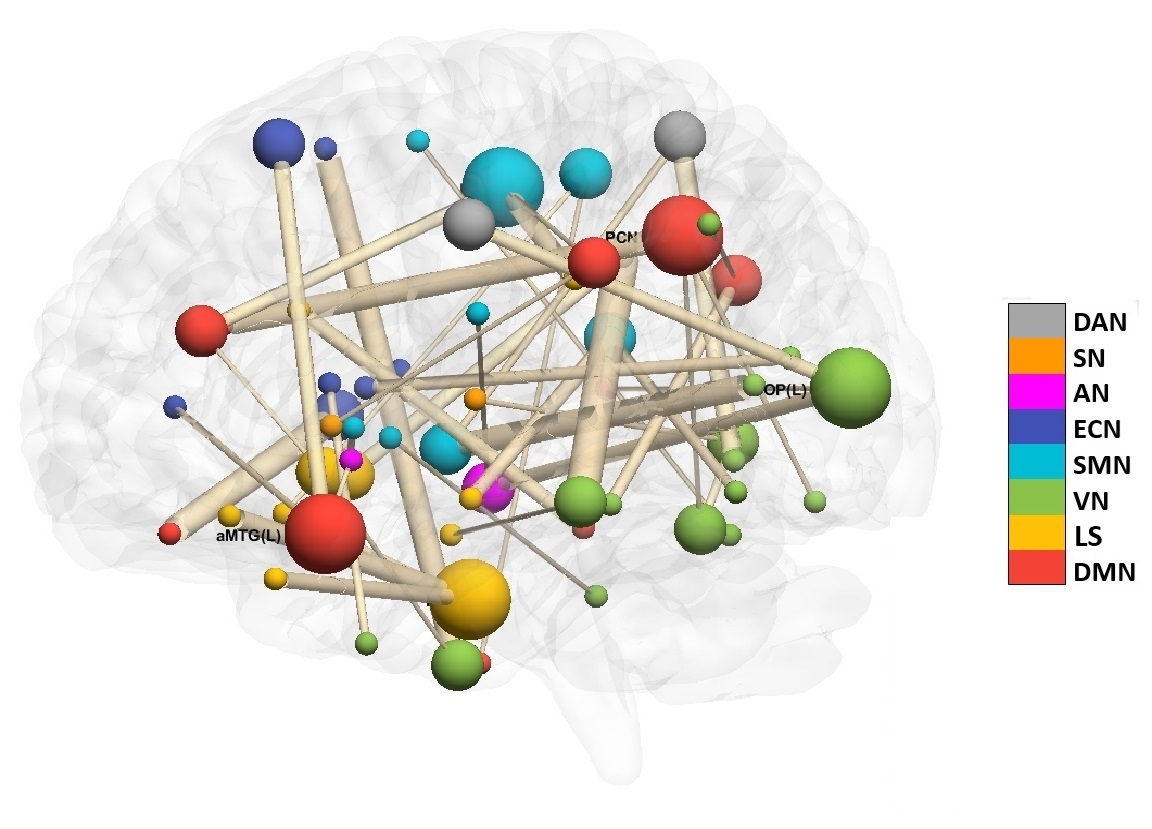}
\caption{All 43 Identified Links}
\label{fig:IP43}
\end{figure}

Of the 206 FC links, only 43 FC mediators exhibited a nonzero IE (see Figure \ref{fig:IP43}). In this figure, nodes represent ROIs, with larger nodes indicating hubs (i.e., ROIs with more connected FC links). The lines between nodes represent FC links, with line thickness reflecting the strength of the IE. Based on the biomarker detection results, the therapeutic application of SSRIs may alleviate the aberrations observed in these identified FC mediators among individuals with IP which exhibit cognitive deficits, potentially reducing their symptoms. The top 10 FC mediators with nonzero IE are presented in Table \ref{tab:Top10}. The ROIs associated with the top identified FC mediators align with previous research findings by demonstrating a pronounced association with the symptoms of IP, notably depression and anxiety.

\begin{table}[h]
\caption{TOP 10 Identified Links}
\label{tab:Top10}
\centering
\begin{adjustbox}{scale=0.65}
\begin{tabular}{@{}cllllcccccccccc@{}} \\\hline
\multirow{2}{*}{Rank} & \multicolumn{2}{c}{ROI 1}& \multicolumn{2}{c}{ROI 2} & \multirow{2}{*}{IE} & \multirow{2}{*}{$\hat{\alpha}$} & \multirow{2}{*}{$\hat{\beta}$} & \multicolumn{3}{c}{IP-SSRI}& \multicolumn{3}{c}{HC} & SSRI vs HC \\ \cline{2-5}\cline{9-14}
 & Name&Network &Name&Network&&& & Pre & Post &Change & Pre &Post &Change & Change \\ \hline
1 & aSMG(L) &DAN& toITG(L)&VN & -1.321 & -0.132 & 10.025 & 0.467 & 0.312 & -0.154 & 0.281 & 0.307 & 0.026 & -0.181 \\
2 & FMC &DMN& IFGtr(R)&ECN & -1.074 & 0.155 & -6.934 & -0.075 & 0.020 & 0.095 & -0.008 & -0.113 & -0.105 & 0.200 \\
3 & TFa(R) &VN& SFG(R) &ECN & -0.978 & 0.139 & -7.059 & -0.125 & 0.052 & 0.177 & 0.036 & 0.036 & 0.000 & 0.177 \\
4 & PAL(R) &SMN& CALC(L)&VN & -0.977 & 0.097 & -10.123 & 0.063 & 0.066 & 0.002 & 0.073 & -0.056 & -0.129 & 0.131 \\
5 & PCN &DMN&PAC(R) &DMN & -0.918 & 0.104 & -8.821 & 0.119 & 0.180 & 0.061 & 0.212 & 0.132 & -0.080 & 0.141 \\
6 & PHa(R)&LS& FOC(L) &LS & -0.788 & -0.155 & 5.072 & 0.165 & 0.033 & -0.132 & 0.001 & 0.067 & 0.066 & -0.198 \\
7 & LING(R) &VN& SPL(R) &DAN & -0.763 & 0.068 & -11.262 & -0.058 & -0.009 & 0.049 & 0.045 & -0.014 & -0.059 & 0.108 \\
8 & PCC&DMN&PRG(R)&SMN & -0.751 & -0.110 & 6.835 & -0.137 & -0.206 & -0.070 & -0.244 & -0.161 & 0.083 & -0.152 \\
9 & PHa(R)&LS&TP(L)&LS & -0.647 & -0.145 & 4.456 & 0.466 & 0.328 & -0.138 & 0.269 & 0.320 & 0.051 & -0.189 \\
10 & aMTG(L)&DMN& SFG(L)&ECN & -0.632 & -0.083 & 7.578 & 0.101 & 0.067 & -0.034 & 0.112 & 0.202 & 0.090 & -0.124 \\ \hline
\end{tabular}
\end{adjustbox}
\end{table}

For instance, the FC mediator with the most substantial IE effect (-1.321) is situated between two specific ROIs: the left anterior division of the supramarginal gyrus (SMG, BA 40) and the left temporooccipital part of the inferior temporal gyrus (toITG, BA 20). SMG, an integral component of the DAN, is implicated in top-down control of attention. Abnormality in SMG has been observed in individuals diagnosed with late-life depression (LLD) or social anxiety disorder (SAD) \citep{jung2018altered,bhaumik2023development}. In addition, the inferior temporal gyrus, a constituent of the visual network, plays a vital role in cognitive processes such as visual processing, object recognition, and semantic processing. Dysfunction in ITG has been identified and demonstrated to be associated with SAD, MDD, and LLD \citep{liu2023childhood, frick2013altered}.

The FC mediator with the second most substantial IE (-1.074) is identified between the frontal medial cortex (FMC) and the right pars triangularis of the inferior frontal gyrus (IFGtr, BA 45). The FMC, part of the DMN, includes the medial prefrontal cortex (mPFC) and ventral medial prefrontal cortex (vmPFC). It plays a crucial role in various cognitive functions, mood regulation, and self-referential tasks. The IFGtr, part of the ECN, is involved in higher-order cognitive processes. Both ROIs are recognized in the literature as critical areas associated with IP symptoms \citep{drevets2008brain, murrough2016reduced, kim2011anxiety, yen2024BLMM, wang2024novel}.

\section{Conclusion and Discussion}
This work enhances biomarker detection in high-dimensional mediation models by introducing novel tuning parameter strategies for $L_1$-norm penalties on DE and IE. The first strategy mitigates the overestimation of DE observed in conventional regularization methods such as Pathway LASSO, thereby enhancing the TPR of IE. The second strategy improves the ability to identify mediators with nonzero $\bm{\beta}$, leading to a better detection of biomarkers. Collectively, these strategies contribute to a robust and accurate analytical framework for navigating the complexities inherent in mediation analysis. Furthermore, we propose an optimal method for selecting the scaling factor within the SIS procedure that significantly boosts the TPR for mediators, especially in scenarios characterized by ultra high-dimensional mediators. Our proposed model is versatile and can be applied to various fields that deal with high-dimensional data, such as genomics, epidemiology, finance, and social sciences. 

The current work applies the tuning parameter strategies to the single-modality mediation model involving FC. Investigating a multimodal approach integrating FC and SC may augment biomarker identification. Furthermore, this work primarily focuses on penalized point estimation, which yields a biased estimate of correlations among mediators. Alternative methodologies, such as debiased LASSO, permutation test, and bootstrapping technique, should be considered to assess the significance of nonzero-effect mediators. In addition, existing penalties including Pathway LASSO do not fully utilize brain network information. To better detect biomarker hubs for neuromodulation and faster recovery, future research could benefit from exploring penalty strategies such as network-constrained penalties \citep{li2008network}.

\backmatter

\section*{DATA AVAILABILITY STATEMENT}
The data is openly available at https://github.com/psyen0824/IPdata.

\bibliographystyle{biom} \bibliography{sample}

\section*{Supporting Information}
Web Appendix A is available with this paper at the Journal website on Wiley Online Library.
\vspace*{-8pt}

\end{document}





\title{Supporting Information for “A Novel Strategy for Detecting Multiple Mediators in High-Dimensional Mediation Models"}
\author{Pei-Shan Yen$^{1}$, Soumya Sahu$^{1}$, Debarghya Nandi$^{1}$, Olusola Ajilore$^{2}$, and\\ Dulal Bhaumik$^{1,2*}$ \\
$^{1}$Division of Epidemiology and Biostatistics, University of Illinois at Chicago, Chicago, US \\
$^{2}$Department of Psychiatry, University of Illinois at Chicago, Chicago, US\\
$^{*}$email: dbhaumik@uic.edu}\\

\appendix
\maketitle
\section{The 105 brain regions of interest}

The brain network was segmented into 105 ROIs based on the CONN atlas \citep{whitfield2012conn}. These ROIs include 50 pairs of bilateral regions and 5 central regions, as detailed in Web~Table~\ref{tab:ROI}. In total, these ROIs account for 5460 unique FC (or links), calculated using the formula for combinations: $\binom{105}{2} = 105 \times 104/2$.

Web~Table~\ref{tab:ROI} also presents the functional networks as defined by \cite{tessitore2019functional}: auditory network (AN), central executive network (CEN), dorsal attention network (DAN), default mode network (DMN), salience network (SN), sensorimotor network (SMN), and visual network (VN). To comprehensively study IP patients, we have included the limbic system (LS) \citep{rajmohan2007limbic}. 
                
\begin{table}[h!]
\fontsize{9}{9} \selectfont 
\caption{The 105 brain regions of interest in CONN atlas}
\centering
\label{tab:ROI}
\begin{threeparttable}
\begin{tabular}{ccllrrr}
\hline
\multirow{2}{*}{Network$^\text{a}$}& \multirow{2}{*}{ROI$^\text{b}$}  & \multirow{2}{*}{  Region Name }& \multirow{2}{*}{Brodmann Area}  & \multicolumn{3}{c}{Coordinates} \\ \cline{5-7}
 &  & && $x$ &$y$ & $z$ \\ \hline
 \multirow{5}{*}{AN} & H & Heschl's Gyrus & 41,43 & -45.2 & -20.3 & 7.2 \\
 & PP & Planum Polare & 38 & -46.6 & -6.0 & -7.3 \\
 & PT & Planum Temporale  & 22 & -52.7 & -29.7 & 10.8 \\
 & aSTG & Superior Temporal Gyrus anterior division & 22,41-42 & -56.2 & -3.9 & -8.0 \\
 & pSTG & Superior Temporal Gyrus posterior division & 22,41-42& -62.3 & -29.2 & 3.8 \\\hline
 \multirow{2}{*}{DAN}& SPL & Superior Parietal Lobule  & 5,7 & -29.3 & -49.5 & 57.5  \\
& aSMG & Supramarginal Gyrus anterior division & 40 & -56.8 & -32.8 & 37.2 \\\hline
 \multirow{8}{*}{DMN} & AG & Angular Gyrus & 39 & -50.4 & -55.7 & 29.7  \\
 & FMC & Frontal Medial Cortex &10-12, 24-25, 32-33& -4.9 & 43.4 & -18.2  \\
& aITG & Inferior Temporal Gyrus anterior division & 20 & -48.1 & -5.0 & -39.2 \\ 
& aMTG & Middle Temporal Gyrus anterior division  & 21 & -57.5 & -4.2 & -22.1 \\
   & PAC & Paracingulate Gyrus &32& -6.2 & 36.7 & 20.8 \\
 & PHp & Parahippocampal Gyrus posterior division & 28, 34-36& -21.9 & -32.4 & -16.9 \\
 & PCN & Precuneous Cortex & 7, 31 & -7.7 & -60.0 & 37.4 \\
 & pSMG & Supramarginal Gyrus posterior division & 40 & -54.9 & -46.0 & 33.2 \\\hline
\multirow{6}{*}{ECN} & CAU & Caudate & -& -12.8 & 9.0 & 9.7 \\
 & FP & Frontal Pole & 10 & -24.7 & 53.0 & 7.5 \\
 & IFGop & Inferior Frontal Gyrus pars opercularis  & 44 & -50.6 & 14.5 & 15.4 \\
 & IFGtr & Inferior Frontal Gyrus pars triangularis & 45 & -49.7 & 28.5 & 8.7 \\
& MFG & Middle Frontal Gyrus & 46 & -38.1 & 18.4 & 42.1 \\
 & SFG & Superior Frontal Gyrus & 8 & -14.1 & 18.7 & 56.2 \\\hline
\multirow{7}{*}{SMN} & CO & Central Opercular Cortex  & 44,47 & -48.0 & -8.6 & 11.8 \\
 & SMA & Juxtapositional Lobule Cortex & 6 & -5.4 & -2.8 & 56.1 \\
 & PAL & Pallidum & 10 & -19.0 & -5.1 & -1.3 \\
 & PO & Parietal Operculum Cortex &40,43 & -48.4 & -31.9 & 20.5 \\
 & poCG & Postcentral Gyrus & 1,2,3 & -38.4 & -27.9 & 51.7 \\
 & PRG & Precentral Gyrus  & 4 & -33.7 & -11.8 & 49.4 \\
 & PUT & Putamen &- & -24.9 & 0.5 & 0.3 \\\hline
\multirow{3}{*}{SN}  & FO & Frontal Operculum Cortex  & 6 & -39.7 & 18.3 & 4.5 \\
 & IC & Insular Cortex & 13-16 & -36.4 & 1.2 & 0.1 \\
& THL & Thalamus &- & -10.2 & -19.3 & 6.3 \\\hline
\multirow{15}{*}{VN} & CN & Cuneal Cortex & 17 & -8.2 & -80.3 & 27.1 \\
& pITG & Inferior Temporal Gyrus posterior division & 20 & -53.4 & -28.5 & -26.0 \\
& toITG & Inferior Temporal Gyrus temporooccipital part & 20 & -51.8 & -53.4 & -16.5  \\
& CALC & Intracalcarine Cortex & 17 & -10.2 & -75.0 & 8.0  \\
 & iLOC & Lateral Occipital Cortex inferior division & 19 & -45.1 & -75.5 & -1.9 \\
 & sLOC & Lateral Occipital Cortex superior division & 19 & -32.0 & -72.9 & 38.0 \\
 & LING & Lingual Gyrus & 17  & -12.3 & -65.7 & -5.4 \\
 & pMTG & Middle Temporal Gyrus posterior division & 21 & -60.9 & -27.4 & -11.0 \\
 & toMTG & Middle Temporal Gyrus temporooccipital part & 21 & -57.6 & -53.0 & 0.8  \\
& OF & Occipital Fusiform Gyrus  & 37 & -26.6 & -76.6 & -13.6  \\
 & OP & Occipital Pole & 17-19 & -16.9 & -96.5 & 6.7 \\
 & SCLC & Supracalcarine Cortex & 17,18& -2.1 & -79.7 & 13.7 \\
 & TFa & Temporal Fusiform Cortex anterior division & 37  & -31.9 & -4.4 & -41.9 \\
 & TFp & Temporal Fusiform Cortex posterior division & 37 & -36.0 & -29.5 & -25.1 \\
& TOF & Temporal Occipital Fusiform Cortex & 37 & -33.5 & -53.7 & -16.0 \\\hline 
\multirow{9}{*}{LS} & Acb & Accumbens  &- & -9.5 & 11.5 & -7.2 \\
 & AMYG & Amygdala &- & -23.0 & -4.9 & -17.7 \\
 & ACC & Cingulate Gyrus anterior division & 24,32 & -4.5 & 17.7 & 24.6 \\
  & PCC & Cingulate Gyrus posterior division & 23,31 &-5.7 & -37.8 & 29.6 \\
 & FOC & Frontal Orbital Cortex & 10,11,47 & -29.5 & 23.7 & -16.6 \\
 & HP & Hippocampus  & -& -25.2 & -23.2 & -13.8 \\
& PHa & Parahippocampal Gyrus anterior division  & 28, 34-36 & -21.9 & -9.1 & -30.3 \\
 & SC & Subcallosal Cortex & 25 & -4.9 & 20.5 & -14.8 \\
 & TP & Temporal Pole & 38 & -40.5 & 11.1 & -29.6 \\\hline
\end{tabular}
\begin{tablenotes} 
\item[a] The abbreviations for brain networks: AN = Auditory Network; CEN = Central Executive Network; DAN = Dorsal Attention Network; DMN = Default Mode Network; SN = Salience Network; SMN = Sensorimotor Network; VN = Visual Network; LS =  Limbic System. 

\item[b] The abbreviations for ROIs follow the standardized terminology of the National Center for Biotechnology Information (NCBI). Brodmann Areas (BA) and stereotactic coordinates (x, y, z) are provided to facilitate anatomical localization and interpretation of the ROIs. For brevity, we present coordinates only for ROIs in the left hemisphere, except for five midline structures: the Frontal Medial Cortex, Subcallosal Cortex, Precuneus Cortex, anterior division of the Cingulate Gyrus, and posterior division of the Cingulate Gyrus.
\end{tablenotes}
\end{threeparttable}
\end{table}

\bibliographystyle{biom} \bibliography{sample}